%% file: main.tex
\documentclass{article}
\usepackage{spconf}
\usepackage{amsmath,amssymb,amsfonts}
\usepackage{graphicx}
\usepackage{hyperref}
\usepackage{cite}
\usepackage{booktabs}
\usepackage{microtype}
\usepackage{url}
\usepackage{textcomp}
\usepackage{xcolor}
\usepackage{multirow}
\usepackage{siunitx}
\usepackage{placeins}
\usepackage{adjustbox}
\usepackage{wrapfig}

\usepackage[nolist]{acronym}
\begin{acronym}
\acro{stft}[STFT]{Short-Time Fourier Transform}
\acro{istft}[iSTFT]{Inverse Short-Time Fourier Transform}
\acro{tf}[T-F]{time-frequency}
\acro{fft}[FFT]{Fast Fourier Transform}
\acro{dft}[DFT]{Discrete Fourier Transform}
\acro{tfrs}[TFRs]{Time-Frequency Representations}
\acro{snr}[SNR]{Signal-to-Noise Ratio}
\acro{rir}[RIR]{room impulse response}
\acro{mse}[MSE]{mean-squared error}
\acro{rmse}[RMSE]{root-mean-squared error}
\end{acronym}

\makeatletter
\g@addto@macro\small{%
  \setlength\abovedisplayskip{6pt plus 2pt minus 1pt}%
  \setlength\belowdisplayskip{6pt plus 2pt minus 1pt}%
  \setlength\abovedisplayshortskip{0pt plus 2pt}%
  \setlength\belowdisplayshortskip{2pt plus 2pt minus 1pt}}
\makeatother
\makeatletter
\def\section{\@startsection{section}{1}{\z@}%
  {-11pt plus -2pt minus -1pt}
  {6pt plus 1pt minus 1pt}
  {\large\bf\fontfamily{ptm}\selectfont\uppercase}}
\def\subsection{\@startsection{subsection}{2}{\z@}%
  {-9pt plus -1pt minus -1pt}
  {6pt plus 1pt minus 1pt}
  {\it\fontfamily{ptm}\selectfont}}
\makeatother

\renewcommand{\paragraph}[1]{\par\smallskip\noindent\textbf{#1}\hspace{0.5em}}

\makeatletter
\g@addto@macro\small{%
  \setlength\abovedisplayskip{4pt plus 2pt minus 1pt}%
  \setlength\belowdisplayskip{4pt plus 2pt minus 1pt}%
  \setlength\abovedisplayshortskip{0pt plus 2pt}%
  \setlength\belowdisplayshortskip{2pt plus 2pt minus 1pt}}
\makeatother
\begin{document}
\ninept

\title{Estimation of Room Impulse Responses from Handclaps}

\name{
Shih-Yu Lai$^{1,2,3}$
\hspace{0.1em}
Kyung Yun Lee$^2$
\hspace{0.15em}
Nils Meyer-Kahlen$^2$
\hspace{0.15em}
Eloi Moliner$^2$
\hspace{0.15em}
Bing-Yu Chen$^1$
\hspace{0.15em}
Vesa Välimäki$^2$
}

\address{
$^1$National Taiwan University, Taipei, Taiwan\,
$^2$Acoustics Lab, DICE, Aalto University, Espoo, Finland\\
$^3$ MoonShine Animation Studio, Taipei, Taiwan}

\maketitle
\begin{abstract}
Handclaps provide an equipment-free excitation for room acoustics, but their unknown and variable source waveform makes room impulse response (RIR) estimation challenging. In this work, we investigate whether RIRs can be estimated directly from handclaps. 
To this end, we introduce an anechoic handclap dataset containing 2,540 claps from 17 participants, designed to capture variability across natural handclaps and different hand configurations.
We first establish the performance attainable when the excitation clap is known using regularized deconvolution, and show that approximating the unknown excitation by windowing the direct sound from the reverberant recording is insufficient.
To estimate the RIR without a known excitation, we propose using the anechoic handclap recordings to train a deep neural network with a supervised regression objective.
Evaluated on a controlled synthetic benchmark, the proposed neural regressor significantly outperforms windowing-based baselines across all instrumental metrics.
Furthermore, we test the proposed method on handclap recordings measured in real acoustic spaces, showing that the inferred RIR spectra are consistent across different handclap measurements taken in the same room location.
These results showcase the feasibility of directly estimating RIRs from natural handclaps without requiring knowledge of the excitation signal.
\end{abstract}


\section{Introduction}

\begin{figure}[t]
    \centering
    \includegraphics[width=\linewidth]{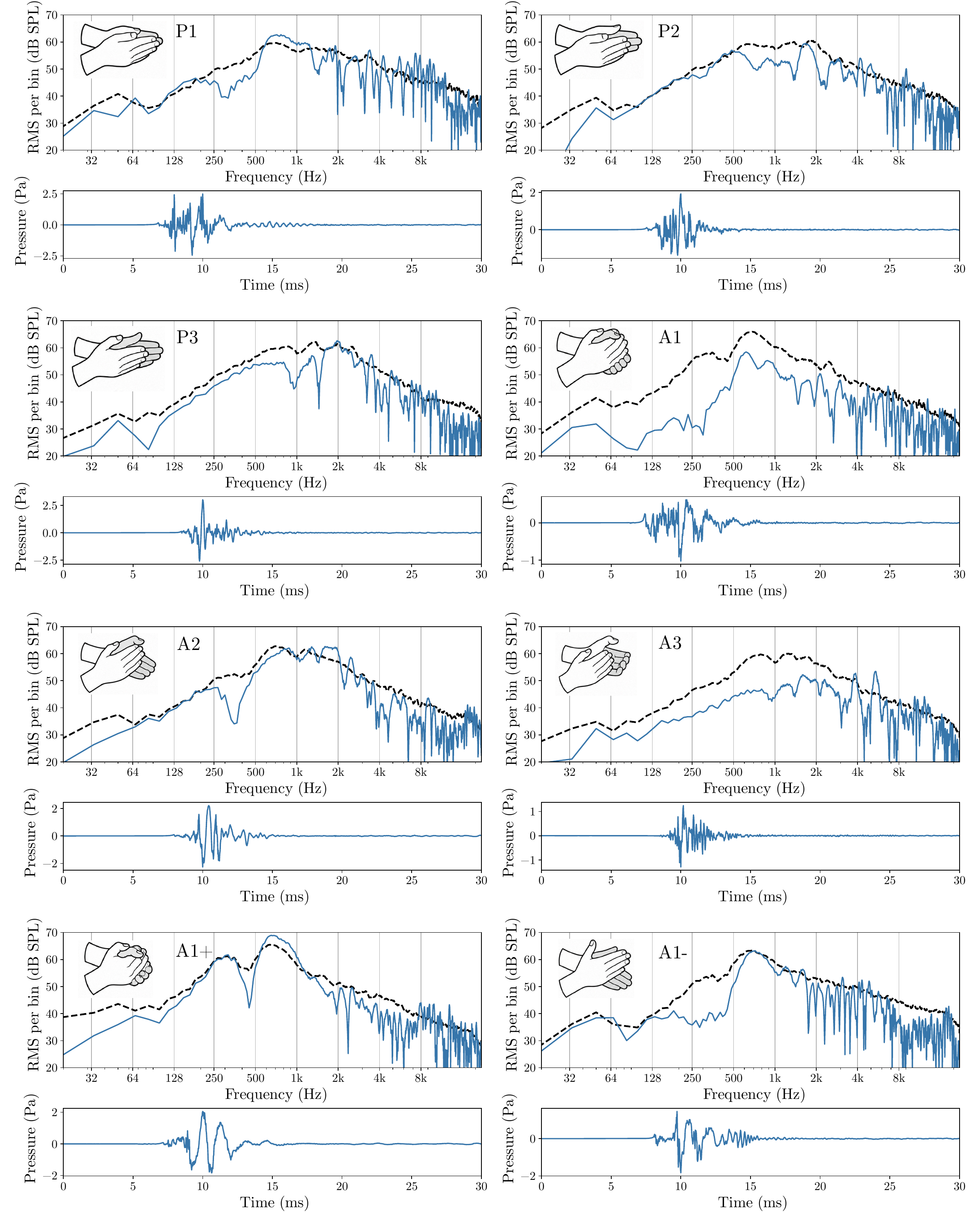}
    \vspace{-0.6cm}
    \caption{Anechoic handclaps of eight different hand posture types. (Dashed line) Mean spectra over entire set of handclaps of that type. (Blue) one example spectrum and corresponding time-domain signal.}
    \label{fig:dataset}
    \vspace{-0.5cm}
\end{figure}

Measuring a \ac{rir} normally requires a known excitation signal reproduced through a loudspeaker at the measurement site.
Established methods, such as exponential sine sweeps \cite{farina2000swept, prawda2022sweep}, provide accurate measurements, but the required equipment and setup limit rapid or opportunistic measurements.
Handclaps offer a convenient alternative because they require no playback hardware, are impulsive, and contain energy over a broad frequency range \cite{Seetharaman2012, papadakis2020handclap, devos2020improved, Fu2025,
halmrast2008simplified,huang2017impulse,rizzi2015rapid}.
However, their spectra are not flat and vary substantially across individuals, hand postures, and even consecutive handclaps by the same person
\cite{repp1987sound,peltola2007synthesis,fletcher2013shock,jylha2008inferring}. Figure\,\ref{fig:dataset} shows examples of claps with different hand postures.

Previous work has explored handclap techniques to improve handclap-based measurements, including gloves to improve the signal-to-noise ratio (SNR) \cite{devos2020improved}.
Handclap responses have been shown to provide useful estimates of reverberation time and energetic room-acoustic parameters \cite{Seetharaman2012, papadakis2020handclap}.
More recently, handclap recordings have also been used to approximate binaural RIRs, and they have been compared with time-stretched pulse measurements \cite{nishimura2025binaural}.

We study the more difficult problem of estimating the \ac{rir} itself from a single recorded handclap without knowing the excitation waveform.
The recorded handclap signal can be modeled as the convolution of an unknown \ac{rir} with the unknown handclap excitation, which entangles excitation-dependent and room-dependent features.
Blindly deconvolving the two factors is particularly ill-posed in this context due to the substantial variability across different handclap executions and individual subjects. Furthermore, we demonstrate that naive approximations, such as windowing the direct path to estimate the source excitation, are insufficient for reliable \ac{rir} recovery.

In this work, we address this problem using a newly recorded, anechoic handclap dataset that captures excitation variability and allows controlled training pairs to be constructed with known \acp{rir}.
Using this dataset, we propose estimating the \ac{rir} from the reverberant handclap using a neural network trained on synthetic paired data with a supervised regression objective.
Experiments on a controlled synthetic benchmark show that the proposed neural regressor significantly outperforms the compared baselines across all acoustic, spectral, and waveform error metrics.


 The remainder of this paper is organized as follows. First, Sec.~\ref{sec:dataset} introduces our publicly available anechoic handclap dataset, capturing natural execution variability across 2,540 handclaps from 17 participants.
 Next, Sec.~\ref{sec:problem} formalizes the handclap deconvolution problem and details our evaluated approaches, including deconvolution with known handclap excitations, estimations of the excitation obtained by windowing the direct path from measurements, and the proposed neural regressor.
 Sec.~\ref{sec:setup} reports both a synthetic benchmark evaluation, built by convolving
simulated and measured 
 \acp{rir} with anechoic handclaps, and a real-world study demonstrating consistent \ac{rir} estimates across different handclaps recorded using a phone.
 Sec.~\ref{sec:conclusion} concludes the paper.
 Code, audio samples, and datasets are available online.\footnote{\url{https://github.com/Akinesia112/ClapRIR.git}}\footnote{\url{https://akinesia112.github.io/ClapRIR/}}



\section{Anechoic Handclap Dataset}
\label{sec:dataset}

\begin{figure}[t]
    \centering
    \includegraphics[width=0.6\linewidth]{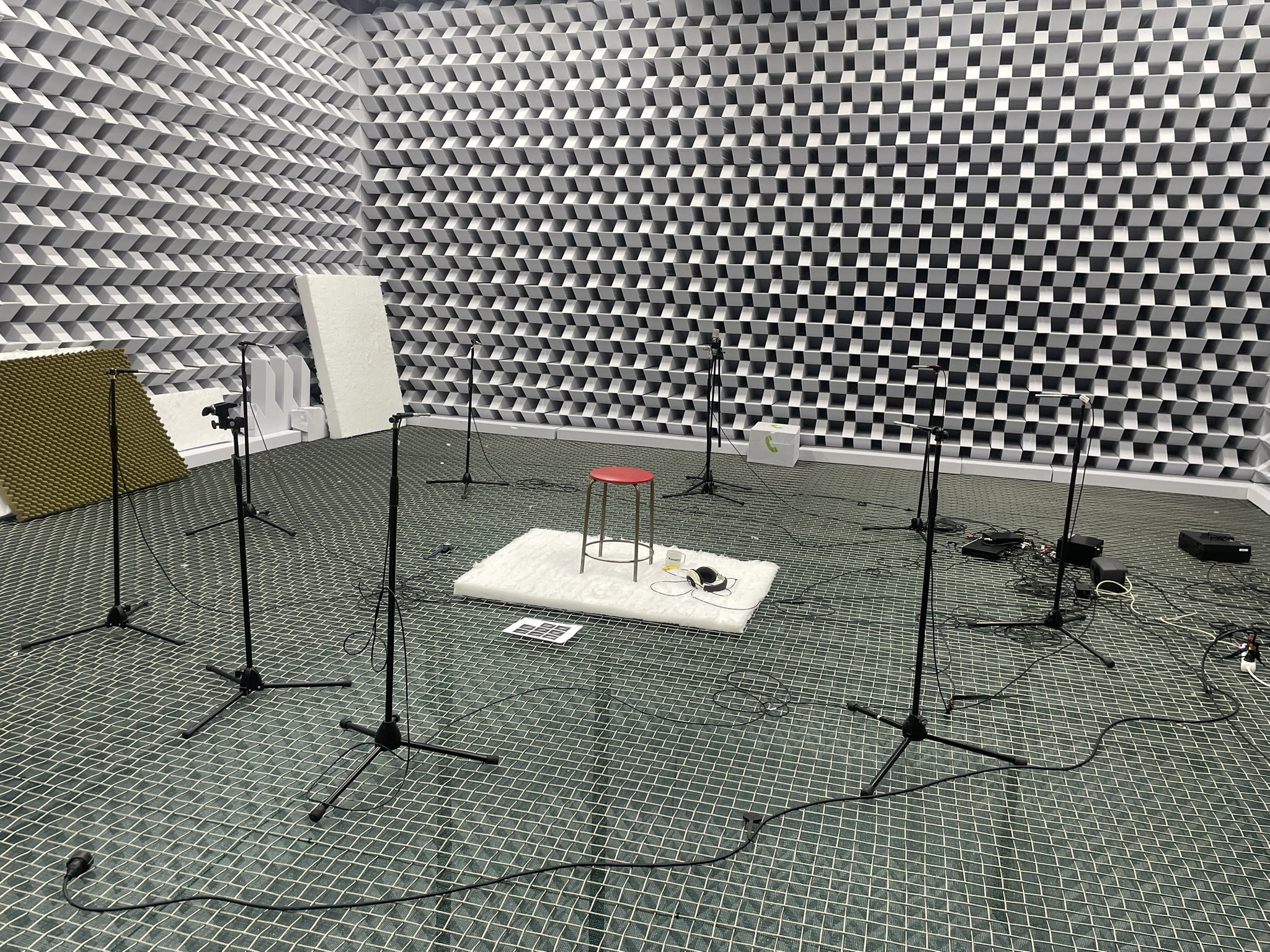}
    \vspace{-0.2cm}
    \caption{Handclap recording setup in the Aalto anechoic chamber.}
    \label{fig:clap_measurement}
    \vspace{-0.4cm}
\end{figure}

We collected isolated handclaps in the large anechoic chamber ``Lampio'' at Aalto University, as shown in Fig. \ref{fig:clap_measurement}.
During recording, 17 participants sat at the center of a circular array of eight equiangularly spaced microphones with a radius of 2~m. Seven of these microphones were G.R.A.S. Type 46AF connected to a G.R.A.S. 12AG preamplifier. The microphone behind the listener belonged to a G.R.A.S. Type 40HF low noise measurement system. All microphones were gain-adjusted after calibration tone-recording using a recording from a BK 4231 calibrator. 

At first, participants were prompted to ``clap as you normally would'' and ``clap as you would when trying to hear the sound of the room''. Then, they were then shown a picture with eight hand configurations from \cite{peltola2007synthesis}. Three positions labeled with ``P1''--``P3'' ask for holding both hands paralel, with decreasing overlap of the hands. The positions labeled with ``A1''--``A3'' follow the same logic, but ask for holding the hands at an angle.
For ``A1$+$'' asks for cupping the hand, and ``A1$-$'' for keeping the fingers straight. In each hand posture, participants were asked to clap 15 times. The protocol therefore requested 150 handclaps per participant. However, some participants produced slightly fewer or additional handclaps.

Recordings were segmented semi-automatically. First, level peaks were identified. Then, handclaps whose maximal level was at least $-30$~dB below the strongest peak of the 15 handclaps of that participant and handclap type were extracted; the extraction windows started 10~ms before and ended 50~ms after each peak. 
Spurious peaks were removed manually after visual inspection.
The final dataset contains 2,540 eight-channel anechoic handclaps and is publicly available.
\footnote{ \url{https://doi.org/10.5281/zenodo.22892926}}

Figure~\ref{fig:dataset} shows average spectra per requested handclap type, along with one example in time and frequency domain each.
All handclaps show bandpass characateristics. 
As analyzed previously, the center frequency depends on the handclap type \cite{peltola2007synthesis}, albeit to a lesser extent, which may be due to stronger inter-individual variability. Yet, there are big differences between individual claps. 
In addition, the examples contain some sharp spectral minima at high frequencies. 
Defining handclap duration by the interval between the two $-60$ dB crossings relative to the peak , the median durations range from 4.85 ms for A3 to 6.46 ms for A1+ (the postures with cupped hands).
\section{Methods}
\label{sec:problem}

Let $h[t]$ denote a \ac{rir}, the signal recorded in a room when the source emits an ideal impulse, and let $y[t]$ denote the signal recorded at the same position when the excitation is a handclap instead, where $t \in \{0, 1, \ldots, T\}$ denotes the discrete-time index.
We model handclap measurements as the convolution of an unknown anechoic handclap $x[t]$ with the unknown \ac{rir} $h[t]$,
\begin{equation}
y[t] = (x \ast h)[t] + n[t],
\label{eq:forward}
\end{equation}
where $\ast$ denotes discrete convolution and $n[t]$ collects measurement noise and other additive perturbations. 
Our goal is to recover the \ac{rir} $h[t]$ from a single handclap measurement $y[t]$.


\subsection{Identifiability and known-excitation reference}
\label{sec:notches}

We first consider the case in which the anechoic handclap excitation $x[t]$ is known. 
Let $Y[k]$, $X[k]$, $H[k]$, and $N[k]$ denote the \acp{dft} of $y[t]$, $x[t]$, $h[t]$, and $n[t]$ respectively, so that \eqref{eq:forward} becomes $Y[k]=X[k]H[k] + N[k]$.
If the measurements are noiseless, $N[k]=0$, a simple deconvolution $H[k] = Y[k]/X[k]$ recovers the RIR exactly wherever $X[k]\neq 0$.
With noise present, the same division yields
\begin{equation}
    \widehat H_{0}[k] =
    H[k]+\frac{N[k]}{X[k]},
    \label{eq:direct_noise}
\end{equation}
so the error term is inversely weighted by the excitation magnitude and grows without bound as $|X[k]|\to0$. Handclaps have deep spectral minima, as seen in Fig.~\ref{fig:dataset}, which can strongly amplify additive perturbations and motivates regularization when such perturbations are present.

Regularization is introduced by replacing the division with a
Tikhonov-regularized deconvolution \cite{kirkeby1999digital},
\begin{equation}
\widehat H_{\lambda}[k]
=
\frac{X^*[k]Y[k]}
{|X[k]|^{2}+\lambda},
\label{eq:tikhonov}
\end{equation}
where $(\cdot)^*$ denotes complex conjugation and $\lambda>0$ is a regularization parameter. 
At bins where $|X[k]|^{2}\gg\lambda$ the estimate reduces to direct division, whereas at the nulls, where $|X[k]|^{2}\ll\lambda$, it is driven toward zero. The parameter thus trades noise amplification against bias.
Thus, when additive perturbations are present, regularization can suppress the large errors caused by division near spectral minima, at the cost of introducing some bias.



\subsection{Windowed-excitation baseline}
\label{sec:excitation}

We now turn to the practical blind scenario in which the excitation $x[t]$ is unknown, and describe a simple baseline.
A natural approximation is to take the excitation from the measurement itself, applying a short initial window that isolates the direct sound,
\begin{equation}
x_{\mathrm{win}}[t]
=
y[t]\, \mathbf{1}[t<T_\mathrm{win}],
\label{eq:crop}
\end{equation}
where $\mathbf{1}[\cdot]$ is the indicator function, equal to one when the
condition holds and zero otherwise, and $T_\mathrm{win}$ is the window
length, set to 3 or 6~ms in our experiments.

Tikhonov-regularized deconvolution \eqref{eq:tikhonov} can then be applied with $x_\mathrm{win}[t]$ in place of the unknown $x[t]$. However, the approximation $x_\mathrm{win}[t] \approx x[t]$ is fundamentally limited. Analysis of the recordings shows that handclaps can last more than 6~ms. 
However, while using a 6~ms window captures more of the handclap, in many rooms it also includes early reflections. A shorter 3-ms window may avoid that, but it truncates the excitation.
We therefore evaluate signals cropped using 3-ms and 6-ms windows, expecting suboptimal performance in both cases. 


\subsection{Neural regressor}

We propose directly regressing the \ac{rir} from a single reverberant handclap, $\mathbf{h}\approx \widehat{\mathbf{h}} = F_{\theta}(\mathbf{y})$, using a deep neural network $F_{\theta}$ trained in a supervised manner.
We write $\mathbf{y}$, $\mathbf{x}$, and $\mathbf{h}$ for the vectors collecting all samples from the reverberant handclap recording, the anechoic handclap excitation, and the \ac{rir}, respectively.
This data-driven formulation replaces the explicit forward model with a mapping learned from examples, so no estimate of the excitation is required.
Supervised training requires paired examples of measurements and their \acp{rir}, which are not available for recorded handclaps. We therefore construct pairs synthetically, drawing anechoic claps $\mathbf{x} \sim p_x$ from our dataset and \acp{rir} $\mathbf{h} \sim p_h$ from a separate dataset of measured RIRs, and convolving them according to \eqref{eq:forward}.
The network is trained by minimizing
\begin{equation}
    \min_{\theta} \;
    \mathbb{E}_{\mathbf{x} \sim p_x, \, \mathbf{h} \sim p_h}
    \left[
        \mathcal{L} \big( \mathbf{h}, \, F_{\theta}(\mathbf{x} \ast \mathbf{h}) \big)
    \right].
    \label{eq:objective}
\end{equation}
The training loss combines a waveform \ac{mse} with an auxiliary magnitude-compressed \ac{stft} loss, written as
\begin{equation}
\mathcal{L}(\mathbf{h},\widehat{\mathbf{h}})
=
\operatorname{MSE}\!\left(\mathbf{h},\widehat{\mathbf{h}}\right)
+
\lambda_{\mathrm{STFT}}
\operatorname{MSE}\!\left(
\Phi_{\alpha}(\mathbf{h}),
\Phi_{\alpha}(\widehat{\mathbf{h}})
\right),
\label{eq:loss}
\end{equation}
where $\Phi_{\alpha}(\mathbf{z})
=
\left(
|\operatorname{STFT}(\mathbf{z})|+\epsilon
\right)^{\alpha}
$.
To design the network $F_\theta$, we use the two-stage time--frequency/time-domain architecture of \cite{moliner2026ambisonics}, adapted to the single-channel case; architecture and training details are given in Sec.~\ref{sec:model_training}.
\section{Experiments and Results}
\label{sec:setup}

\subsection{Training and testing data setup}
\label{sec:data}

Reverberant handclaps are synthesized by convolving anechoic claps with measured \acp{rir} from MIT, BUT, ACE, and OpenAIR datasets \cite{mit,but,ace,OpenAIR}, and also simulated shoebox \acp{rir} simulated with pyroomacoustics \cite{scheibler2018pyroomacoustics}.
The simulated \acp{rir}, as well as those from MIT, BUT, and ACE are used for training with room-disjoint splits, while OpenAIR is reserved for testing.
Up to 16 distinct \acp{rir} are selected from each room in each split, and
each selected \ac{rir} is convolved with five handclaps drawn from the
participants of the corresponding split.
All \acp{rir} are resampled to 44.1~kHz, trimmed to start at their onset, truncated or zero-padded to 1~s, and peak-normalized.
Each \ac{rir} is convolved with five handclaps drawn from the participants of the corresponding split, and each observation is peak-normalized. No additional observation noise is added; measurement-chain effects in measured \acp{rir} remain part of the target response.

We use handclaps from 12 participants for training and 3 for testing (P08, P15, and P17).\footnote{Due to additional, automatic quality control, only 2,065 handclaps are currently used for training, where two participants were excluded completely.} Training handclaps are reduced to 20 ms. Also, we used the samples from all handclap types for our experiments. 
The resulting test set contains 432 clap--\ac{rir} observations, comprising 216 measured and 216 simulated \acp{rir}.

\subsection{Model architecture and training details} \label{sec:model_training}

For $F_\theta$, we reimplement the hybrid architecture of \cite{moliner2026ambisonics}, originally proposed for Ambisonics encoding of multichannel \acp{rir}, and adapt it to a single-channel input and output: the network takes a 1-s reverberant handclap and outputs the corresponding 1-s \ac{rir}.
The first stage computes an \ac{stft} with a 510-sample Hann window and a hop of 128 samples.
 A 2D convolutional NCSN++ network \cite{song2021sde} (25.0M parameters) processes the real and imaginary parts of this complex \ac{stft}, and an inverse \ac{stft} of its output gives a coarse estimate of the full response.
The second stage consists of a 1D U-Net (1.5M parameters) operating on the waveform, which refines the split early reflections (using a mixing time of $\sim$70~ms).
Its output is added to the output of the first stage.

The loss in \eqref{eq:loss} uses $\lambda_{\mathrm{STFT}}=0.25$, $\alpha=2/3$, and $\epsilon=10^{-8}$, with the STFT computed using a 510-sample Hann window and a 127-sample hop.
A mask is applied to both loss terms so the zero-padding applied to measured \acp{rir} which files had been cropped shorter than 1\,s does not contribute to the gradient. 
We train for 20,000 iterations using AdamW \cite{loshchilov2019adamw} with a constant learning rate of $2\times10^{-4}$, gradient-norm clipping at 1, and an effective batch size of 16. 
Full training details are available in the accompanying code.

\subsection{Evaluation metrics}
\label{sec:metrics}

We report four reference-based metrics comparing the estimated \ac{rir} $\widehat{\mathbf{h}}$ with the corresponding reference $\mathbf{h}$.
The first one is the normalized energy-decay-convergence (EDC)~\cite{dal2024similarity} computed broadband, which compares the Schroeder energy decay of the estimated \ac{rir} $\widetilde{D}_{\widehat{\mathbf{h}}}$ with the corresponding reference $\widetilde{D}_{\mathbf{h}}$. The metric is reported in dB as:
\begin{equation}
\mathcal{L}_{\mathrm{EDC}}^{\mathrm{dB}}
=
10\log_{10}\!\left(
\operatorname{MSE}\!\left(
\widetilde{D}_{\widehat{\mathbf{h}}},
\widetilde{D}_{\mathbf{h}}
\right)
/
\operatorname{mean}\!\left(
\widetilde{D}_{\mathbf{h}}^{\,2}
\right)
\right).
\label{eq:edc}
\end{equation}
Because each decay curve is referred to its own initial level, $\mathcal{L}_{\mathrm{EDC}}^{\mathrm{dB}}$ is invariant to overall gain and measures normalized broadband decay-shape error. 
We also report the absolute early decay time (EDT) error, estimated from the Schroeder energy decays and reported as $\Delta \mathrm{EDT}=|\mathrm{EDT}(\widetilde{D}_{\widehat{\mathbf{h}}})-\mathrm{EDT}(\widetilde{D}_{\mathbf{h}})|$ in milliseconds.

The third metric is the log-spectral error (LSE), defined as the mean squared difference between the log-magnitude \acp{stft} of the estimated and reference \acp{rir}:
\begin{equation}
\mathcal{L}_{\mathrm{LSE}}
=
\operatorname{mean}\!\left(
\left(
\log \left| \mathbf{X}_{\widehat{\mathbf{h}}} \right| 
- 
\log \left| \mathbf{X}_{\mathbf{h}} \right|
\right)^2
\right),
\label{eq:lse}
\end{equation}
where $\mathbf{X}_{\widehat{\mathbf{h}}}$ and $\mathbf{X}_{\mathbf{h}}$ are the complex time-frequency \ac{stft} representations of $\widehat{\mathbf{h}}$ and $\mathbf{h}$, respectively.
Finally, we report the normalized root-mean-squared error (NRMSE) in the waveform domain:
\begin{equation}
\mathcal{L}_{\mathrm{NRMSE}}
=
\sqrt{
\operatorname{MSE}\!\left(
\widehat{\mathbf{h}}, \mathbf{h}
\right)
/\operatorname{mean}\!\left(
\mathbf{h}^2
\right)
}
.
\label{eq:nrmse}
\end{equation}
Lower values indicate better performance for all metrics.



\subsection{RIR estimation evaluation}
\label{sec:results}


We compare the neural regressor against the windowed-excitation baseline (Sec.~\ref{sec:excitation}) using window lengths $T_\mathrm{win}$ of 3\,ms and 6\,ms. 
To show how well the \ac{rir} can be recovered under ideal conditions when the anechoic handclap is known, we evaluate both unregularized and Tikhonov-regularized deconvolution (Sec.~\ref{sec:notches}). 


Table~\ref{tab:table1} reports the results separately for the measured and simulated test \acp{rir}.
Under known-excitation conditions, Tikhonov regularization improves the measured-RIR results, reducing NRMSE from $0.061$ to $0.022$, whereas unregularized deconvolution is slightly better for simulated \acp{rir} ($0.060$ vs.\ $0.069$).
Thus, regularization is beneficial for the measured-RIR block in this benchmark, but is not universally required.
When the excitation is unknown, both windowed-excitation baselines yield substantially higher estimation errors than the neural regressor, showing a significant performance gap relative to the known-excitation reference. 
These results illustrate the fundamental limitation of direct-sound windowing: a short window ($T_\mathrm{win}=3\,\text{ms}$) truncates the excitation signal, whereas a longer window ($T_\mathrm{win}=6\,\text{ms}$) incorporates early room reflections that severely degrade the subsequent deconvolution. While both windowing strategies yield substantial estimation errors, the 6-ms window degrades performance even further than the 3-ms variant.

\input{tables/main_results_nostd_onlymeasured}



The neural regressor improves upon windowed-excitation baselines across all reported metrics, for both measured and simulated \acp{rir}.
On the measured set, for instance, $\mathcal{L}_\mathrm{LSE}$ drops from
$0.20$ to $0.112$ and $\mathcal{L}_\mathrm{NRMSE}$ from $2.7$ to $0.733$
relative to the best-performing windowed baseline.
However, a performance gap relative to the known-excitation reference remains.


\subsection{RIR estimation consistency on measured handclaps}
\label{sec:real}

To evaluate the performance of the proposed method on a practical real-world setup, we recorded a total of 280 handclaps in 7 different rooms.
For each room, 40 handclaps were performed, 5 for each of the positions introduced in Sec.~\ref{sec:dataset}.
Recordings were captured with an iPhone 15 at a fixed source-receiver distance of approximately 2~m, resampled to 44.1~kHz, and time-aligned using the recorded onset pre-pad, without further preprocessing.
We analyze whether the \ac{rir} estimates generated by the neural regressor remain consistent across individual handclap measurements within the same environment.

\begin{figure}[t]
    \centering
    \vspace{-0.9cm}
   \includegraphics[width = 0.98\columnwidth]
    {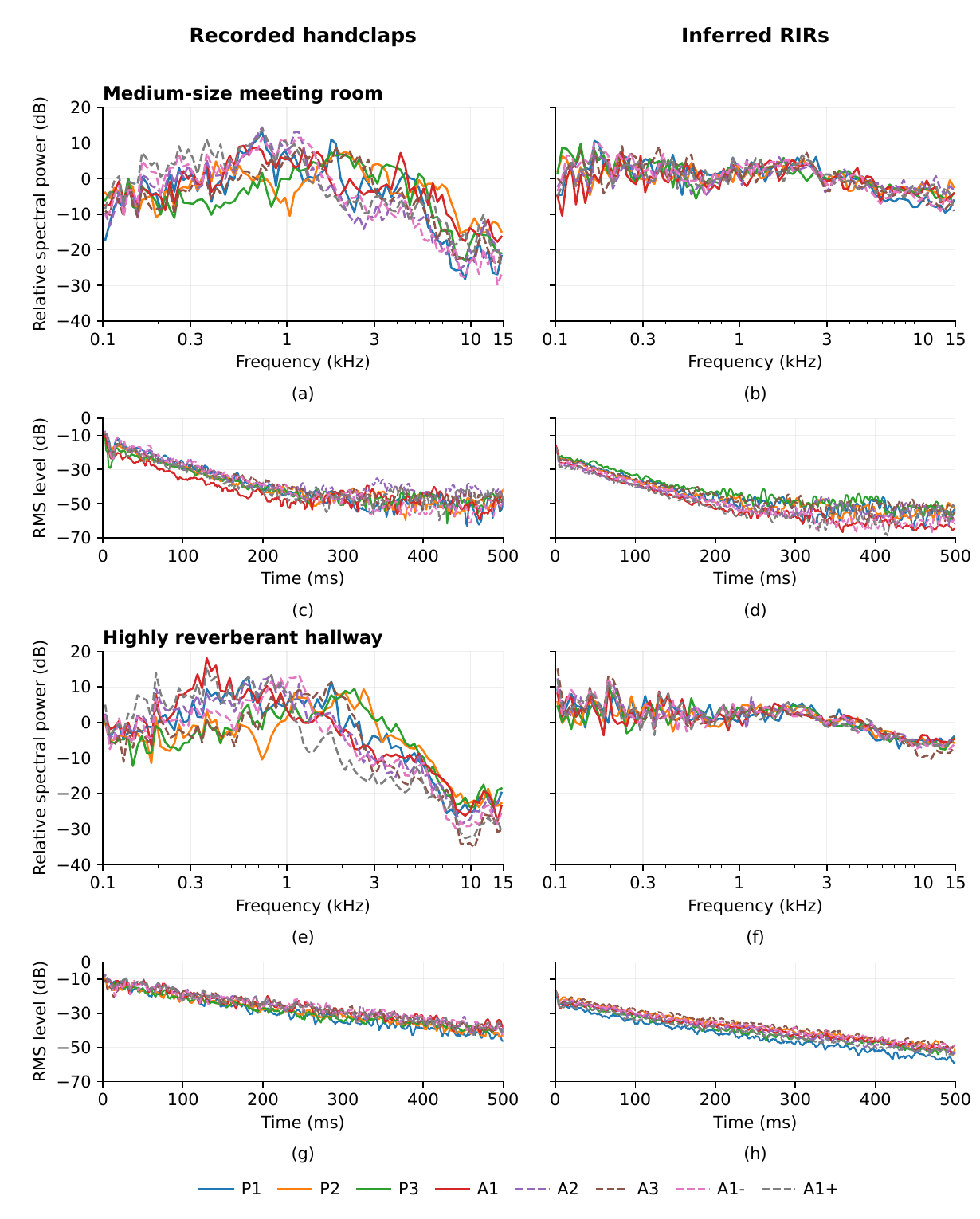}
    \vspace{-0.5cm}
\caption{Phone-recorded handclaps (left) and corresponding \ac{rir} estimates obtained with the neural regressor (right) for eight handclap positions in two rooms, showing spectra and RMS envelopes}
\label{fig:phone-spectrum}
    \vspace{-0.5cm}
\end{figure}

Fig.~\ref{fig:phone-spectrum} shows qualitative estimation results for two representative environments, showing 8 handclap measurements per room.
The upper panels (a), (b), (e), and (f) show 1/12-octave-averaged spectra over 0.1--15~kHz using a common 0--0.9-s interval, with each spectrum normalized to its mean FFT-bin power over 0.1--10~kHz.
The lower panels (c), (d), (g), and (h) show RMS envelopes computed with a 5-ms centered window and an approximately 1-ms hop over the first 500~ms on a common digital-amplitude scale in dB, without per-signal normalization.
The spectra of the estimated \acp{rir} are visibly more consistent across positions than those of the raw handclaps, which exhibit distinct handclap-dependent resonances. Additionally, the RMS envelopes demonstrate a consistent decay profile within each room, closely matching the energy decay of the handclap recordings.

To quantify the consistency observed in Fig.~\ref{fig:phone-spectrum}, we compute the spectral dispersion across all measurements within each room.
Let $S_i(b)$ denote the log-magnitude spectrum (in dB) of the $i$-th handclap across $F=87$ twelfth-octave bands. 
For a given room, the spectral dispersion $D_r$
is defined as the root-mean-square spectral distance  averaged over all $\binom{40}{2}=780$ unique pairs among $M=40$ hanclaps:
\begin{equation}
D_r
=
\frac{1}{\binom{M}{2}}
\sum_{i<j} 
\sqrt{
\frac{1}{F}
\sum_{b=1}^{F}
\left(S_i(b)-S_j(b)\right)^2
}.
\label{eq:spectral-dispersion}
\end{equation}
Across the 7 rooms, the neural regressor reduces the mean spectral dispersion from 6.45\,dB for the raw handclap recordings to 3.43\,dB. In contrast, the 3-ms and 6-ms windowed-excitation baselines yield 6.76\,dB and 6.12\,dB, respectively.
\section{Conclusion}
\label{sec:conclusion}

We studied end-to-end \ac{rir} estimation from handclap recordings. 
 To capture the natural acoustic variations of handclap excitations, we presented a novel dataset of anechoic handclaps. 
We have shown that handclaps contain enough spectral information so that, if the exact excitation signal were known, regularized deconvolution is effective. However, in the strictly blind case, simple methods based on estimating the excitation signal by windowing the measurements fall short.
To address this, we proposed training a neural regressor using our handclap dataset, which outperforms windowed-excitation deconvolution and establishes a robust blind deconvolution baseline. 
A real-world test demonstrates consistent \ac{rir} estimates across different handclap recordings in the same room, indicating that the model generalizes beyond synthetic data and effectively decouples the \ac{rir} from the spectral variations of individual handclaps. 

Although training on measured \acp{rir} promotes generalization, these tarpgets inevitably include the transfer functions of the measurement hardware. As a result, the network regresses toward an average loudspeaker and microphone response rather than isolating the pure room acoustics.
Explicitly decoupling the device responses from target \acp{rir} remains a key direction for future work.
Furthermore, results still indicate a performance gap against the known-excitation deconvolution reference, leaving clear scope for further improvement, such as leveraging multiple handclap recordings to promote a more robust \ac{rir} estimation.
This study demonstrates the potential of deep neural networks to turn casual handclaps into reliable signals for practical room acoustic measurements.





\section{Acknowledgments}
This study was conducted when the first author visited the Aalto Acoustics Lab in Jul.--Sept. 2026. This work was supported by the HUCE infrastructure of the Aalto School of Electrical Engineering, the National Science and Technology Council (NSTC), Taiwan (under NTSC 114-2221-E-002-218-MY3, and 114-2218-E-002-006), and National Taiwan University (114L900902 and 115L8909) funded through the Ministry of Education (MOE), Taiwan. The authors thank Samuel Belzner for providing the iPhone handclap recordings and pictures found on the website; Gloria Dal Santo and Sebastian J. Schlecht for research ideation and metrics design; and the computational resources from the Aalto Science-IT project, MoonShine Animation Studio and National Taiwan University.
Apart from text editing, generative artificial intelligence has been used for generating the handclap pictograms in Fig.\ref{fig:dataset} (Open AI, GPT 6 Atra Light), website and code developement (Claude, Opus 5).

\bibliographystyle{IEEEbib}
\bibliography{2026_claps}

\end{document}

%% file: tables/main_results_nostd_onlymeasured.tex
\begin{table}[t]
\vspace{-7pt}
\caption{Mean \ac{rir} estimation results for both measured and simulated \acp{rir}. 
Known-excitation rows are references; best unknown-excitation result per metric and block is shown in bold.}
\label{tab:table1}

\resizebox{\columnwidth}{!}{%
\centering
\begin{tabular}{@{}ll|rrrr@{}}
\toprule
& \textbf{Method}
& $\mathcal{L}_\mathrm{EDC}^\mathrm{dB}$ $\downarrow$
& {$\Delta\mathrm{EDT}$} $\downarrow$
& $\mathcal{L}_\mathrm{LSE}$ $\downarrow$
& $\mathcal{L}_\mathrm{NRMSE}$ $\downarrow$ \\
\midrule \midrule

\multirow{5}{*}{\textit{Measured}}
& Known exc. (unreg.)
& $-30.8$\,dB
& 8.7\,ms
& 0.071
& 0.061 \\

& Known exc. (Tikhonov)
& $-41.2$\,dB
& 1.8\,ms
& 0.031
& 0.022 \\
\cmidrule{2-6}

& Windowed exc. (3 ms)
& $-18.3$\,dB
& 93\,ms
& 0.20
& 2.7 \\

& Windowed exc. (6 ms)
& $-17.4$\,dB
& 126\,ms
& 0.27
& 3.0 \\

& Neural regressor
& $\mathbf{-22.2}$\,dB
& \textbf{56.9}\,ms
& \textbf{0.112}
& \textbf{0.733} \\
\midrule \midrule

\multirow{5}{*}{\textit{Simulated}}
& Known exc. (unreg.)
& $-52.0$\,dB
& 0.6\,ms
& 0.0035
& 0.060 \\

& Known exc. (Tikhonov)
& $-49.7$\,dB
& 0.8\,ms
& 0.0043
& 0.069 \\
\cmidrule{2-6}

& Windowed exc. (3 ms)
& $-16.7$\,dB
& 26\,ms
& 0.46
& 3.8 \\

& Windowed exc. (6 ms)
& $-15.4$\,dB
& 39\,ms
& 0.51
& 4.9 \\

& Neural regressor
& $\mathbf{-25.3}$\,dB
& \textbf{17.7}\,ms
& \textbf{0.061}
& \textbf{0.79} \\
\bottomrule
\end{tabular}%
}
\vspace{-0.5cm}
\end{table}